\documentclass[twocolumn,showpacs,preprintnumbers,amsmath,amssymb]{revtex4}

\usepackage{graphicx}% Include figure files
\usepackage{dcolumn}% Align table columns on decimal point
\usepackage{bm}% bold math
\input{epsf.tex}

\begin{document}

\title{Observation of Narrow $N^+(1685)$ and $N^0(1685)$ Resonances in
  $\gamma N \to \pi \eta N$ Reactions}

\author{V.~Kuznetsov$^{1,2,*}$}
\author{F.~Mammoliti$^{2,3}$}
\author{F.~Tortorici$^{2,3}$}
\author{V.~Bellini$^{2,3}$}
\author{V.~Brio$^{2,3}$}
\author{A.~Gridnev$^{1}$}
\author{N.~Kozlenko$^{1}$}
\author{G.~Russo$^{2,3}$}
\author{A.~Spatafora$^{2,3}$}
\author{M.~L.~Sperduto$^{2,3}$}
\author{V.~Sumachev$^{1}$}
\author{C.~Sutera$^{2}$}
\vspace{0.5 cm}

\affiliation{$^1$ Petersburg Nuclear Physics Institute, Gatchina 188300, Russia}
\affiliation{$^{2}$INFN - Sezione di Catania, via Santa Sofia 64, I-95123 Catania, Italy}
\affiliation{$^{3}$Dipartimento di Fisica ed Astronomia, Universit\`a di Catania, via Santa Sofia 64, I-95123 Catania, Italy}

%\vspace{0.5 cm}

\begin{abstract}
\noindent 
Observation of a narrow structure at $W\sim 1.68$ GeV in the excitation
functions of some photon- and pion-induced reactions may signal a new narrow
isospin-1/2 $N(1685)$ resonance.  New data on the $\gamma N \to \pi \eta N$
reactions from GRAAL seems to reveal the signals of both $N^+(1685)$ and
$N^0(1685)$ resonances.

\pacs{14.20.Gk,13.60.Rj,13.60.Le}
\end{abstract}
\thanks{Electronic address: Kuznetsov\_va@pnpi.nrcki.ru}
\maketitle

Understanding the internal structure of the nucleon is a key task in the domain
of hadronic physics.  Suggested in the 60th the approximate flavor SU(3)
symmetry of QCD led to a remarkably successful classification of low-lying
mesons and baryons. Many properties of baryons known at that time were
transparently explained by the Constituent Quark Model (CQM)~\cite{cqm} that treats
baryons as bound systems of three effective (constituent) quarks.
% in the ground or excited states.  For decades CQM was accepted as the most
% successful tool for the classification and interpretation of the hadron
% spectrum.
 
CQM-based calculations predicted a rich spectrum of baryon resonances with
widths varying from $\sim 80$ to $\sim 400$ MeV.  Nevertheless, in spite of
significant efforts, many of the predicted resonances still escape from
reliable experimental identification (the so-called ``missing resonances'').

The Chiral Soliton Model ($\chi SM$) is an alternative picture of baryons.  It
treats them as space/flavor rotational excitations of a classical object - a
soliton of the chiral field.  The model predicts the lowest-mass baryon
multiplets to be the $({\bf 8}, 1/2^+)$ octet and the $({\bf 10}, 3/2^+)$
decuplet - exactly as CQM does.  $\chi SM$ also predicts the existence of
long-lived exotic particles~\cite{dia}.

Therefore the search for light-quark exotic states may provide critical
benchmarks to examine two different approaches and to establish the connection
between them.  In this context the observation of a narrow enhancement at $W\sim
1.68$ GeV in the $\gamma n \to \eta n$ excitation function (the so-called
``neutron anomaly'') at GRAAL, CBELSA/TAPS, LNS and A2@MAMI C
~\cite{gra2,kru1,kru2,kas,kru4} might be quite important.  Narrow structures at
the same energy were also observed in Compton scattering on the neutron $\gamma
n \to \gamma n$~\cite{comp} and in the beam asymmetry for the $\eta$
photoproduction of the proton $\gamma p \to \eta p$~\cite{acta} (see alsocite{ann}).
The recent data on the beam asymmetry for Compton scattering on the proton $\gamma p \to
\gamma p$~\cite{comp1}, the precise data for the $\gamma n \to \eta
n$~\cite{wert2} and $\pi^- p \to \pi^- p$~\cite{epe} reactions revealed two
narrow structures at $W \sim 1.68$ and $W \sim 1.72$ GeV.

The whole complex of experimental observations may signal the existence of one
($N(1685)$) or two ($N(1685)$ and $N(1726)$) narrow nucleon resonances. The
properties of $N(1685)$ (if it does exist), namely the isospin 1/2, strangeness
$S=0$, narrow ($\Gamma \leq 25$ MeV) width, strong photoexcitation on the
neutron and suppressed decay to $\pi N$ final state, do coincide well with those
predicted by $\chi SM$ for the second member of the anti-decuplet of exotic
particles~\cite{az}.

On the other hand there are alternative interpretations of the ''neutron
anomaly'' in terms of of the specific interference of known wide
resonances~\cite{ani} or as the sub-threshold meson-nucleon production
(cusp)~\cite{dor}. Although being questionable~\cite{kuz}, the first assumption is widely
discussed in literature.

The decisive identification of these experimental findings is a challenge for
both theory and experiment.  In the previous experiments the possible signal of
$N(1685)$ was observed in so-called ''formation'' reactions in which the
incoming particle interacts with the target nucleon and excites resonances.  If
$N(1685)$ does really exist, its signal should also be seen in multi-particle
''production'' reactions in which it would manifest itself as a peak in the
invariant mass spectra of the final-state products.  Possible reactions could be
$\gamma N \to \pi \eta N$.

The photoproduction of $\pi \eta$ pairs on the proton was previously studied at
GRAAL~\cite{hou}, CBELSA/TAPS~\cite{gutz} and A2@MAMI C~\cite{a2}
facilities. The goals were to investigate the spectrum of baryon resonances and
to constrain theoretical models.  The works~\cite{hou,gutz} were restricted to
only the $\gamma p \to \pi^0 \eta p$ reaction.  The data from Ref.~\cite{a2}
were obtained at photon energies below $1.4$ GeV.

In this Letter, we report on the study of the $\gamma p \to \pi^0 \eta p, \gamma
p \to \pi^+ \eta n, \gamma n \to \pi^0 \eta n$, and $\gamma n \to \pi^-\eta p$
reactions.  Our ultimate goal is to search for a possible signal of $N(1685)$.

The data were collected at the GRAAL facility~\cite{gra_fac}. The GRAAL
highly-polarized beam was produced by means of the back-scattering of laser
light on $6.04$ GeV electrons circulating the storage ring of the European
Synchrotron Radiation Facility (Grenoble, France). The GRAAL tagging system
provided the measurement of photon energies in the range 0.55 - 1.5 GeV. The
maximum beam intensity and polarization were in the energy range 1.4 - 1.5 GeV.

Photons from $\eta \to 2\gamma$ and $\pi^0 \to 2\gamma$ decays were detected in
the BGO Ball~\cite{bgo}. This detector covered the range of polar angles
$\theta_{lab} = 25 - 165^{\circ}$. It made possible to determine the photon
energy with resolution $\frac{3\%}{E_{\gamma}(GeV)}$. The angular resolution of
photon detection was $6 - 8 ^{\circ}$.

\begin{figure} 
\vspace*{-1. cm}
\begin{center} %\resizebox{0.6\columnwidth}{!}{\includegraphics{compcb.eps}}
\epsfverbosetrue\epsfxsize=9.2cm\epsfysize=8.7cm\epsfbox{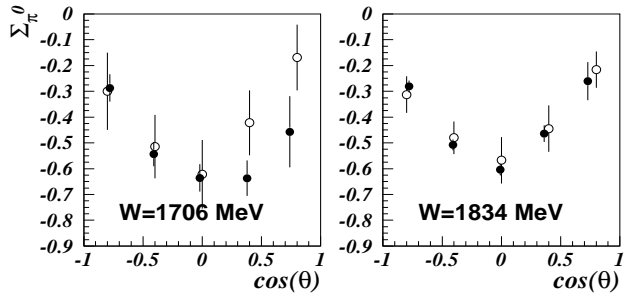}
\vspace*{-4.5 cm}
\caption{Beam asymmetry $\Sigma_{\pi^0}$ obtained assuming the $\gamma p \to
\pi^0 R_{(\eta p)}$ reaction.  Black circles are the results of this work. Open
circles are the results of the CBELSA/TAPS Collaboration~\protect\cite{gutz} }.
\end{center} 
\vspace*{-1.0 cm}
\label{fig1}
\end{figure}

The recoil protons and neutrons emitted at forward angles $\theta_{lab} \leq
25^{\circ}$ were detected in the assembly of forward detectors. It consisted of
two planar wire chambers, a thin scintillator hodoscope and a lead-scintillator
wall~\cite{rw}. Two latter detectors were located at $3$ m far from the
target. They allowed a measurement of time-of-flights of recoiled nucleons with
resolution $\sigma_{TOF}\sim 250$ psec. Then this quantity was used to retrieve
the energies of protons and neutrons. The planar chambers made possible to
measure proton angles with resolution better than $1^{\circ}$. The neutron
angles were measured by the lead-scintillator wall which provided about
$2-3^{\circ}$ resolution.

Charged pions were detected in the BGO Ball. Their angular quantities were
measured by two cylindrical wire chambers which surrounded the target and
provided an angular resolution $1-2^{\circ}$.  The pion energies were reconstructed
assuming the momentum conservation.
 
At the first step of the data analysis $\eta$ and $\pi^0$ mesons were identified
by means of the invariant masses of two properly chosen photons.  Then the cuts
on the proton and $\eta$ missing masses were applied.

The kinematics of a three-body $\gamma N \to \pi \eta N$ reaction could be
considered as a combination of three two-body reactions with one ''real" and one
''effective" two-particle  in the final state.

\begin{eqnarray} \gamma N \to \pi R_(\eta N) \to \pi \eta N \\ \gamma N \to \eta
R_(\pi N) \to \pi \eta N \\ \gamma N \to R_(\pi \eta) N \to \pi \eta N
\end{eqnarray}

At the second stage of the data analysis the cuts on the coplanarity and on the
differences between the missing and invariant masses assuming all three possible
kinematic cases were employed. For the selection of $\gamma N \to \pi \eta N$
events the $\pi$ azimuthal angle was compared with the azimuthal angle of the
''effective'' $\eta N$ resonance $R(\eta p)$ (Eq.~1). The missing mass of
$R(\eta p)$ $MM(\gamma, \pi)$ calculated from the momentum of the incoming
photon and the final-state $\pi$ was compared with the invariant mass of the
final-state $\eta$ and $N$ $IM(\eta N)$. Then similar cuts were imposed assuming two other
reactions (Eq.~2 and 3).

\begin{figure} 
%\vspace*{1. cm}
\begin{center}
\epsfverbosetrue\epsfxsize=4.2cm\epsfysize=3.7cm\epsfbox{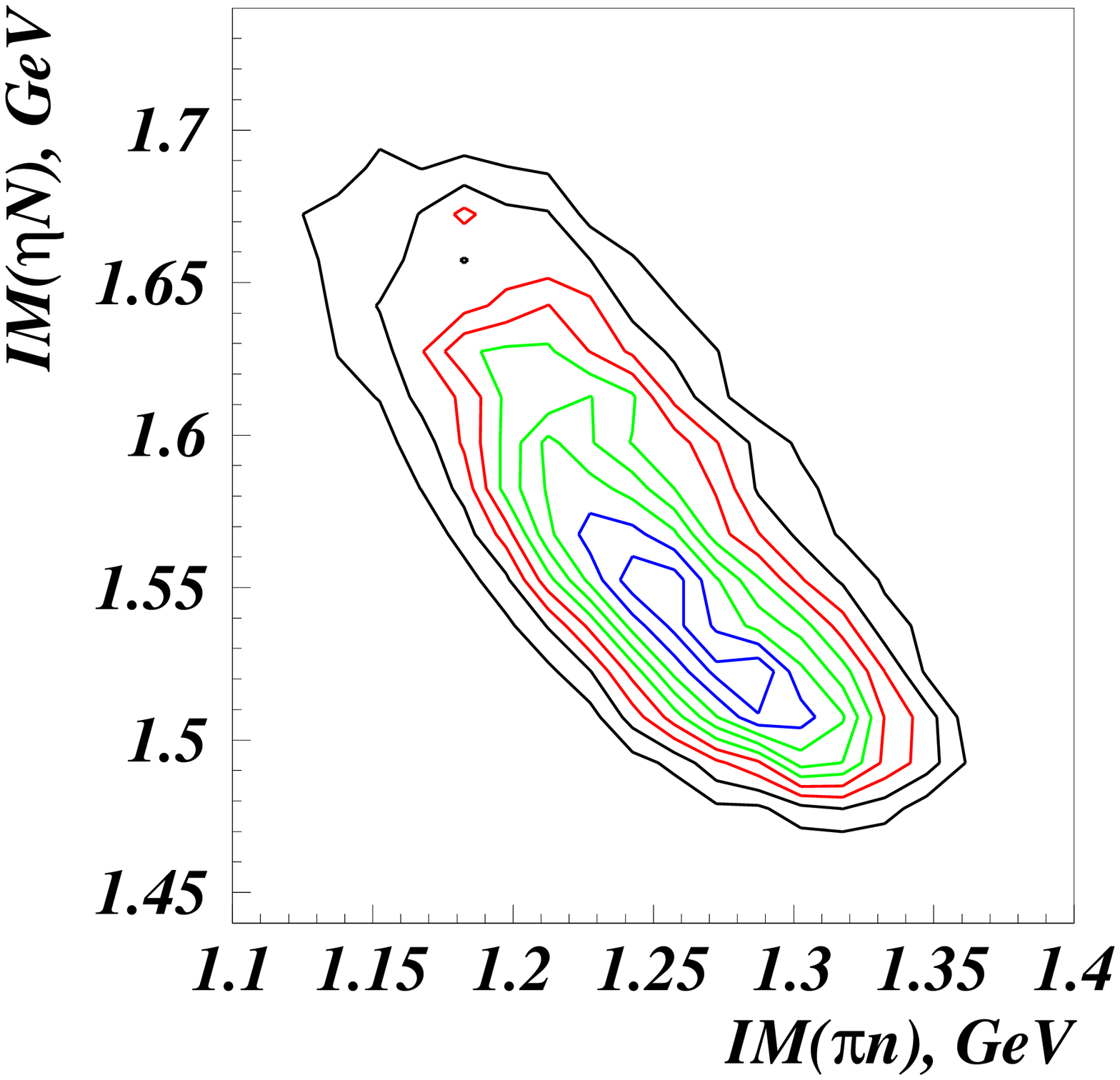}
\epsfverbosetrue\epsfxsize=4.2cm\epsfysize=3.7cm\epsfbox{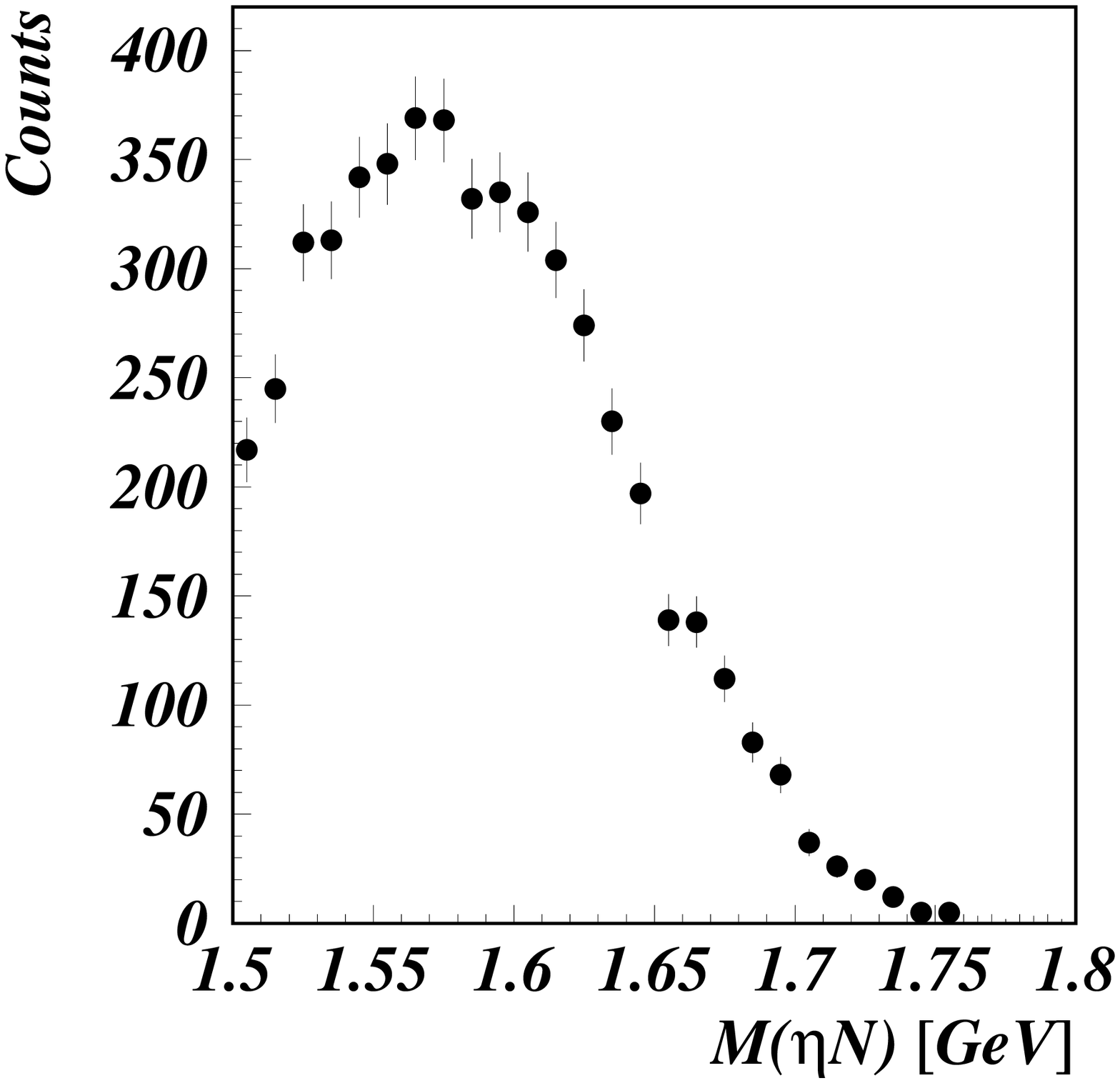}
%\vspace*{-0.5 cm}
\caption{On the left: Bi-dimensional plot of $\eta N$ invariant mass $IM(\eta N)$ versus $\pi N$ invariant
mass $IM(\pi N)$ in the energy range $1.4 - 1.5$ GeV (sum of all reactions under study). 
On the right: the corresponding spectrum of the extracted $M(\eta N)$ mass. Neither of cut 
on the $IM(\pi N)$ was applied.}
\end{center}
\vspace*{-0.5 cm}
\label{fig2}
\end{figure}

These cuts eliminated the background which was clearly seen in the initial
spectra (for example, in the spectra of $\eta$ and proton missing masses).  To
assure the quality of the analysis, our results on the polarized photon beam
asymmetry $\Sigma$ for $\gamma p \to \pi^0 \eta p$ were compared with those
published by the CBELSA/TAPS Collaboration~\cite{gutz} (Fig.1).
Both data sets are in good agreement while our data are
more precise at the energies $E_{\gamma} = 1.2 - 1.45$ GeV (the average
center-of-mass energy $W\sim 1.834$ GeV).

Given the goal of this work, only the events in the range of the energy of the
incoming photon $E_{\gamma}=1.4 - 1.5$ GeV were selected for further analysis.  The lower
limit of $1.4$ GeV is close to the $\gamma N \to \pi N(1685)$ threshold. The
upper value $1.5$ GeV is the limit of the GRAAL beam and it also allows to avoid
the contribution from higher-lying resonances.

The left panel of Fig.2 shows the Dalitz plot of the invariant mass $IM(\eta N)$
versus the invariant mass $IM(\pi N)$ (the sum of all reactions under study).
The events corresponding to $IM(\pi N) \sim 1.2 - 1.35$ GeV are major
contributors.  One may assume that they originate from the $\gamma N \to \eta
\Delta$ production.  There is a small narrow enhancement at $IM(\eta N)\sim
1.68$ GeV.  This enhancement may signal $N(1685)$.  The correspondinf spectrum of the 
extracted invariant mass $M(\eta N)$ is shown on the right panel of Fig.2.

To eliminate the contamination of $\gamma N \to \eta \Delta$ events, further the cuts on the
invariant mass $1.12 \leq IM(\pi N) \leq 1.22$ GeV and the missing mass $MM(\gamma, \eta) \leq 1.22$
GeV were applied to compromise between the
overall statistics of selected events and the rejection of the background.

The spectra of the extracted masses $M(\eta N)$ for each reaction are shown in
Fig.3. For the reaction $\gamma p \to \pi^0 \eta p$ on the
free-proton target $M(\eta p)$ was taken as $(MM(\gamma, \pi^0)+ IM(\eta p))/2$. 
This made possible the most proper usage of the information read out
from the GRAAL detector and consequently to improve the resolution.
In the case of the reaction $\gamma p \to \pi^+ \eta n$ on the free proton 
the energy of the $\pi^+$ is retrieved from the momentum conservation.
That is why in addition a kinematic fit was employed to achieve the best
resolution in the $M(\eta n)$ spectrum.  
For the reactions on the proton and the neutron bound in a deuteron target, the
missing masses $MM(\gamma, \pi)$ are distorted due to Fermi motion of the target
nucleon while the invariant masses $IM(\eta N)$ remain almost unaffected.  For these
reactions the extracted masses $M(\eta N)$ shown in Fig.3 were set equal to the
corresponding invariant masses $IM(\eta N)$.

All the spectra exhibit enhancements at $M(\eta N) \sim 1.68$ GeV. It is better
pronounced and more narrow for the reactions on the free proton (two upper
panels). The statistics for these reactions is better because of the available
data. In the case of the reactions of the proton and neutron bound in the
deuteron the peaks are wider.

A signal of $N(1685)$ resonance should be seen in both missing
mass $MM(\gamma, \pi)$ and invariant mass $IM(\eta N)$.  The left panel of Fig.4 shows a
bi-dimensional plot of these quantities (the sum of all the reactions under study). There is
a clear enhancement at $\sim 1.68$ GeV at both axis. The corresponding spectrum
of the $\eta N$ mass reveals a peak-like structure
at $W\sim 1.68$ GeV. Being considered in conjunction with high-statistics results on
the $\gamma n \to \eta n$~\cite{gra2,kru1,kru2,kas,kru4} and other reactionscite{comp,acta,comp1,wert2,epe},
this structure signals the existence of the $N^(1685)$ resonance.
 
The positions of the peaks in both missing masses $MM(\gamma, \pi)$ and
invariant masses $IM(\eta N)$ for each reaction depend on the quality of the
calibration of the GRAAL sub-detectors and tagging system. This in particular
concerns the forward time-of-flight detectors.  An error in the determination of
the time-of-flight of the recoil nucleon of $\sim 20$ psec results in a shift
of the peak position $\sim 10$ MeV. These errors might be different for recoil protons and recoil neutrons.
Other errors originate from the calibration of
the tagging system $\Delta E_{\gamma}=10$ MeV, from the threshold effects in the BGO Ball, and from the
energy losses of the protons during their passes from the target to the
detectors.  That is why all the spectra were corrected such that the peaks were
located at the same average value $M(\eta N) \sim 1.68$ GeV. The deviation of
the initial peak positions from this average value did not exceeded 10 MeV.

\begin{figure} 
%\vspace*{1.2 cm}
\begin{center}
\epsfverbosetrue\epsfxsize=9.2cm\epsfysize=6.7cm\epsfbox{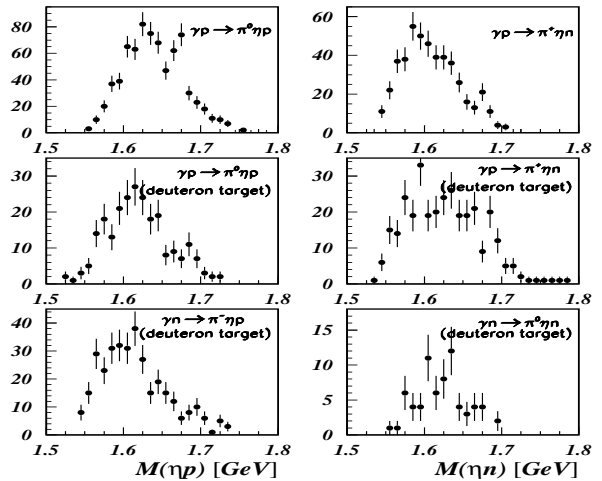}
%\vspace*{-5.0 cm}
\caption{Spectra of extracted masses $M(\eta N)$}.
\end{center} 
\vspace*{-1.5 cm}
\label{fig3}
\end{figure}

\begin{figure} 
\vspace*{1.0 cm}
\begin{center}
\epsfverbosetrue\epsfxsize=4.2cm\epsfysize=3.7cm\epsfbox{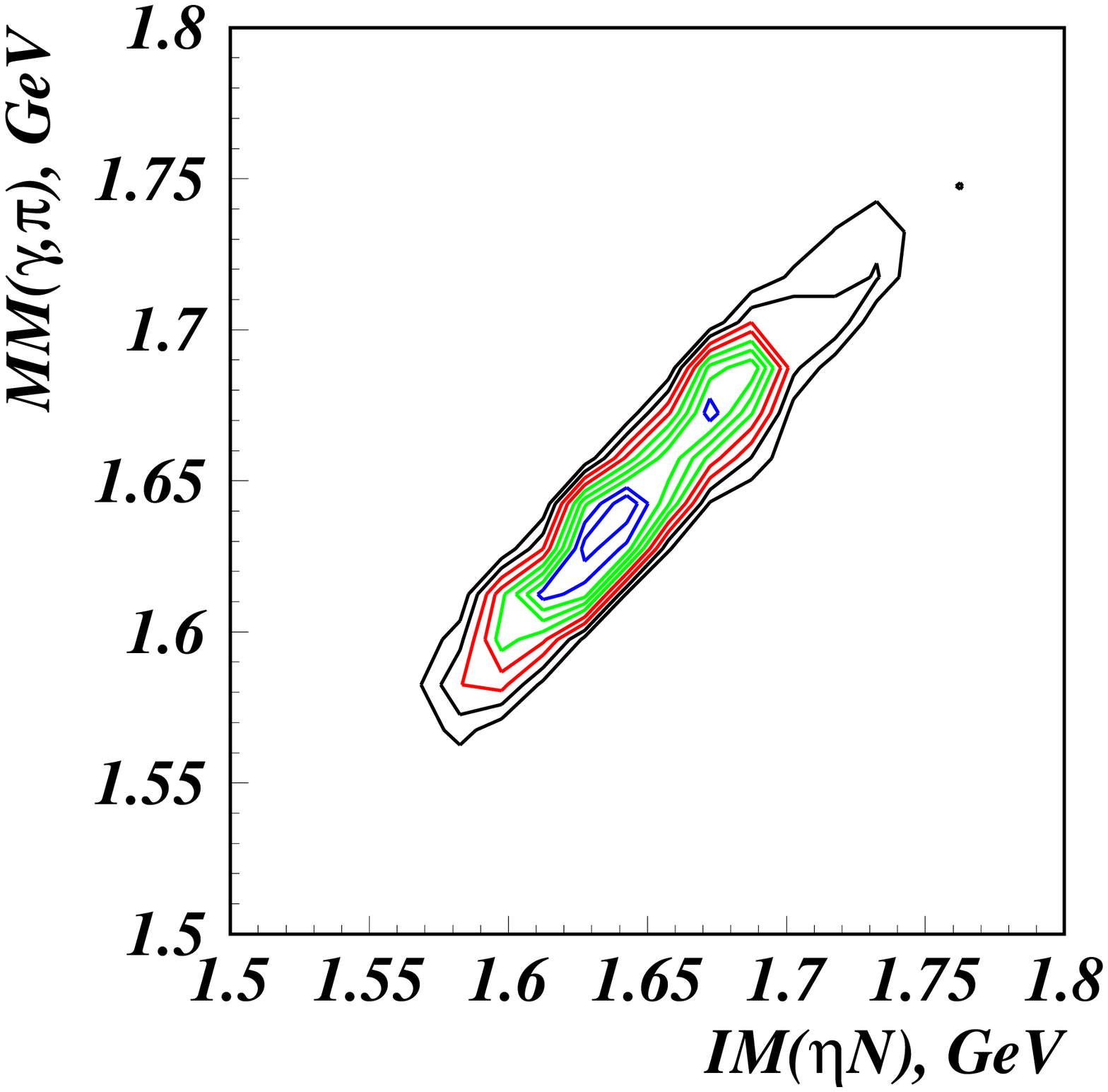}
\epsfverbosetrue\epsfxsize=4.2cm\epsfysize=3.7cm\epsfbox{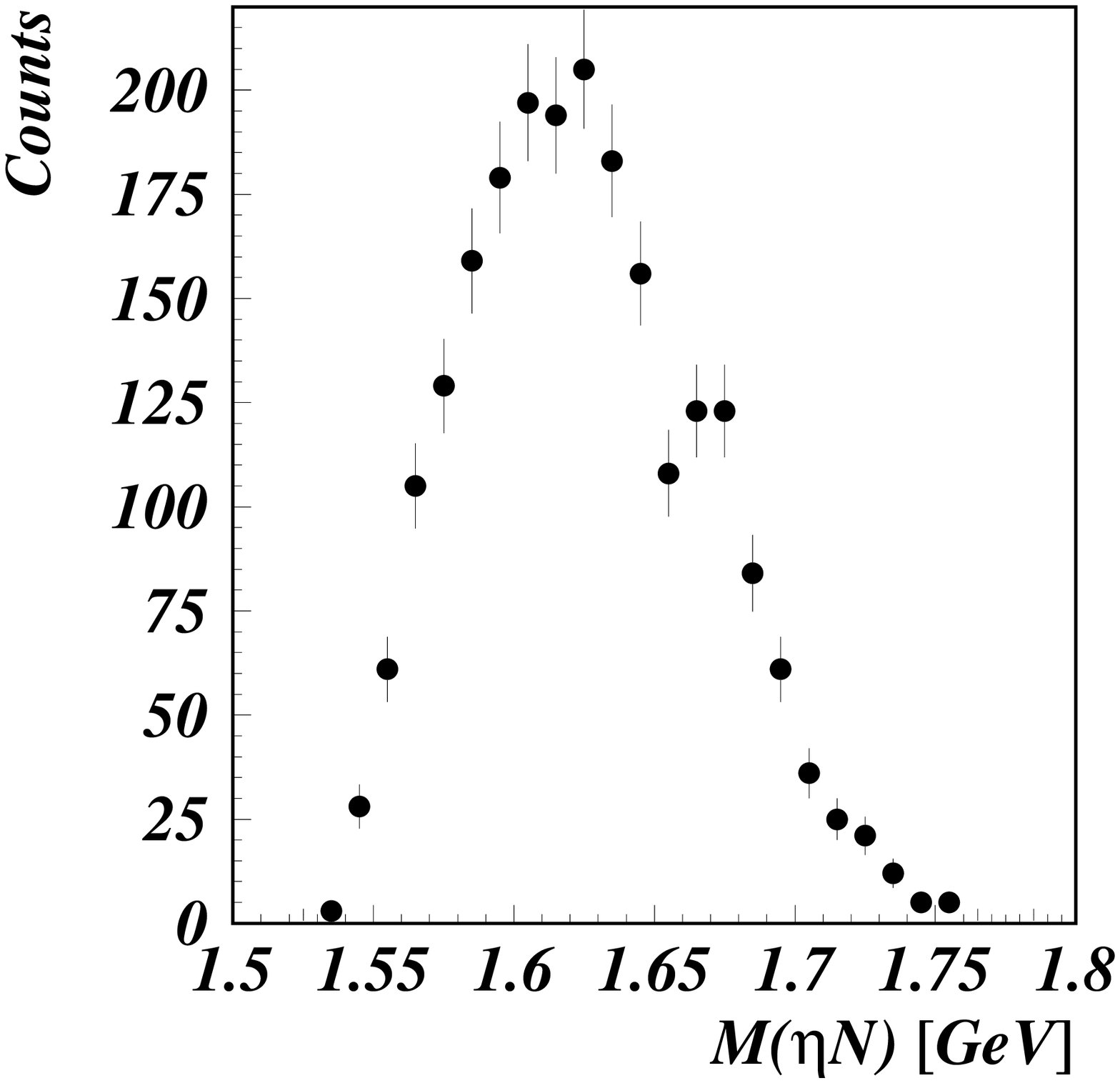}
\protect\vspace*{-2.0 cm}
\vspace*{2.0 cm}
\caption{On the left: Bi-dimensional plot of missing masses $MM(\gamma, \pi)$ vs invariant
masses $IM(\eta N)$ (Sum of all channels) with the cut on $IM(\pi N)$.  On the right: the corresponding spectrum of
the extracted $(M\eta N)$. }
\end{center} 
\vspace*{-0.5 cm}
\label{fig4}
\end{figure}

\begin{figure} 
\vspace*{-0.5 cm}
\begin{center}
\epsfverbosetrue\epsfxsize=6.7cm\epsfysize=6.0cm\epsfbox{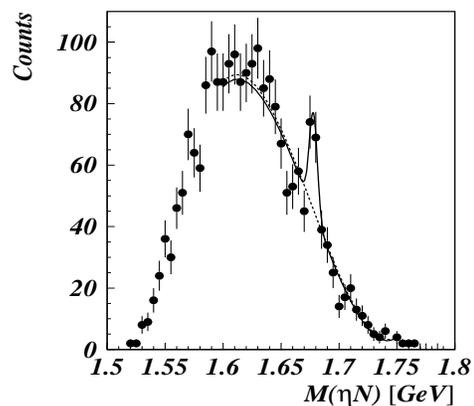}
\vspace*{-0.5 cm}
\caption{Spectrum of extracted $M(\eta N)$ mass (sum of all channels) with corrections (see text for detail).}
\end{center}
\vspace*{-0.3 cm}
\label{fig5}
\end{figure}

The sum of the corrected $M(\eta N)$ spectra is shown in Fig.5.
There is a well pronounced peak at $\sim 1.68$ GeV. The Gaussian+3-order polynomial
(signal-plus-background) fit results in the $\chi$-square of $23.9/23$. The fit
by 3-order polynomial (background) gives the $\chi$-square of $42.6/26$.  The
log likelihood ratio of these two hypotheses ($\sqrt{2\ln (L_{B+S}/L_B)}$)
corresponds to the confidence level of $4.6 \sigma$.

The extracted peak position is $M=1678\pm 0.8_{stat}\pm 10_{syst}$~MeV.  The
systematic uncertainty in the mass position originates from the uncertainties in
the calibration of the GRAAL detector and tagger.  The width $\Gamma \sim 10$
MeV may be affected by the mentioned above corrections.  

Fig. 6 presents the simulated yields of $\gamma p \to \pi^0 \eta p$ and $\gamma p \to \pi^+ \eta n$
events obtained by using the same software and cuts as those for real data shown in the Fig.3.
The event generator used in $MC$ included flat cross sections without any narrow resonances.  Neither of
peaks appeared in the $M(\eta,N)$ spectra.
 
Our results support the existence of two narrow resonances, $N^+(1685)$
decaying, in particular, into $\eta p$ final state, and $N^0(1685)$ with one
possible decay into $\eta n$ (i.e. the isospin-1/2 $N(1685)$ resonance).
Although the properties of this resonance (if it does exist) do coincide well to those expected for
the second member of the exotic anti-decuplet~\cite{dia}, its decisive
accusation requires in particular the identification of the second structure at
$W\sim 1.726$ GeV.

\begin{figure} 
%\vspace*{-0.5 cm}
\begin{center}
\epsfverbosetrue\epsfxsize=4.2cm\epsfysize=3.7cm\epsfbox{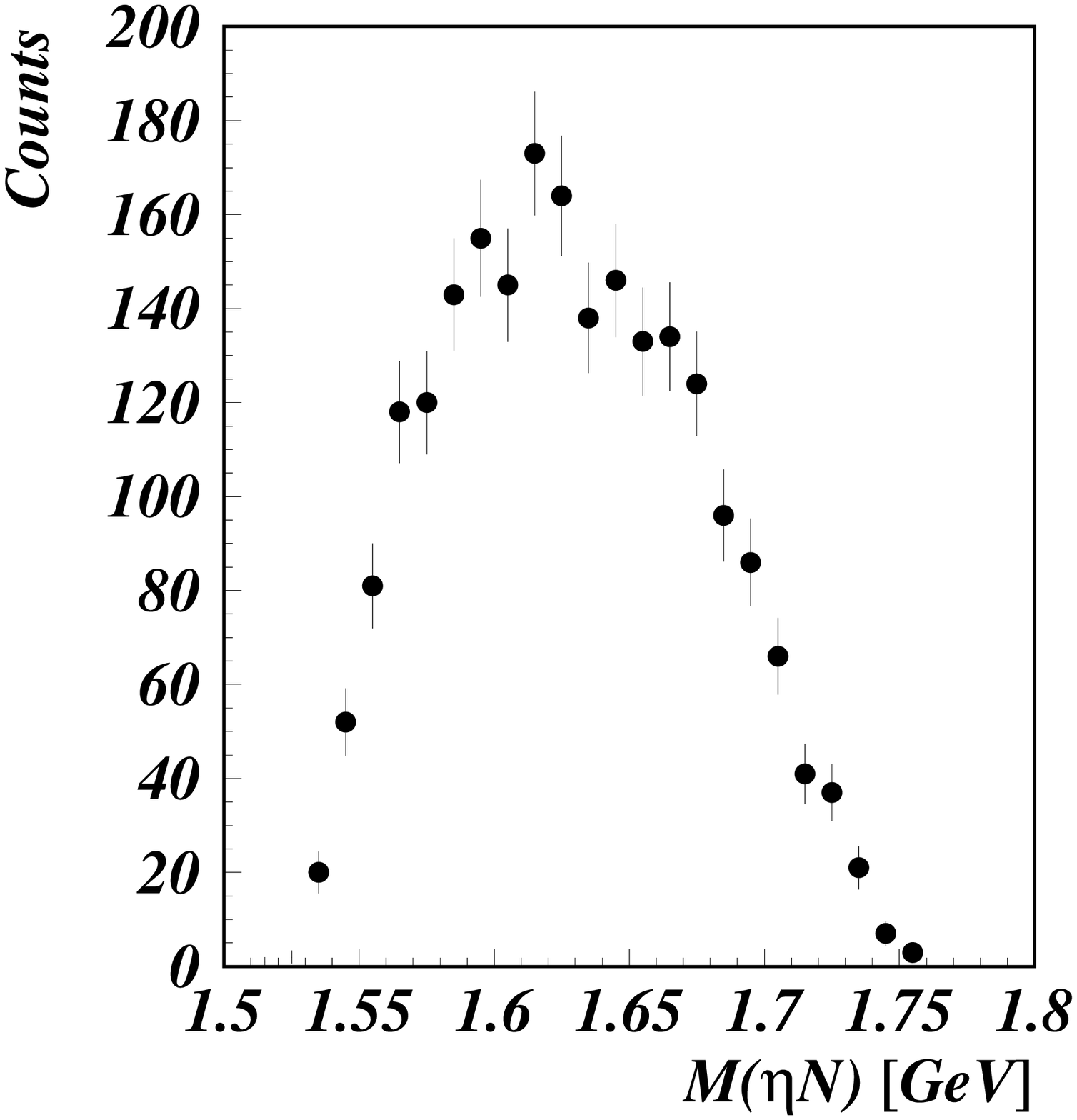}
\epsfverbosetrue\epsfxsize=4.2cm\epsfysize=3.7cm\epsfbox{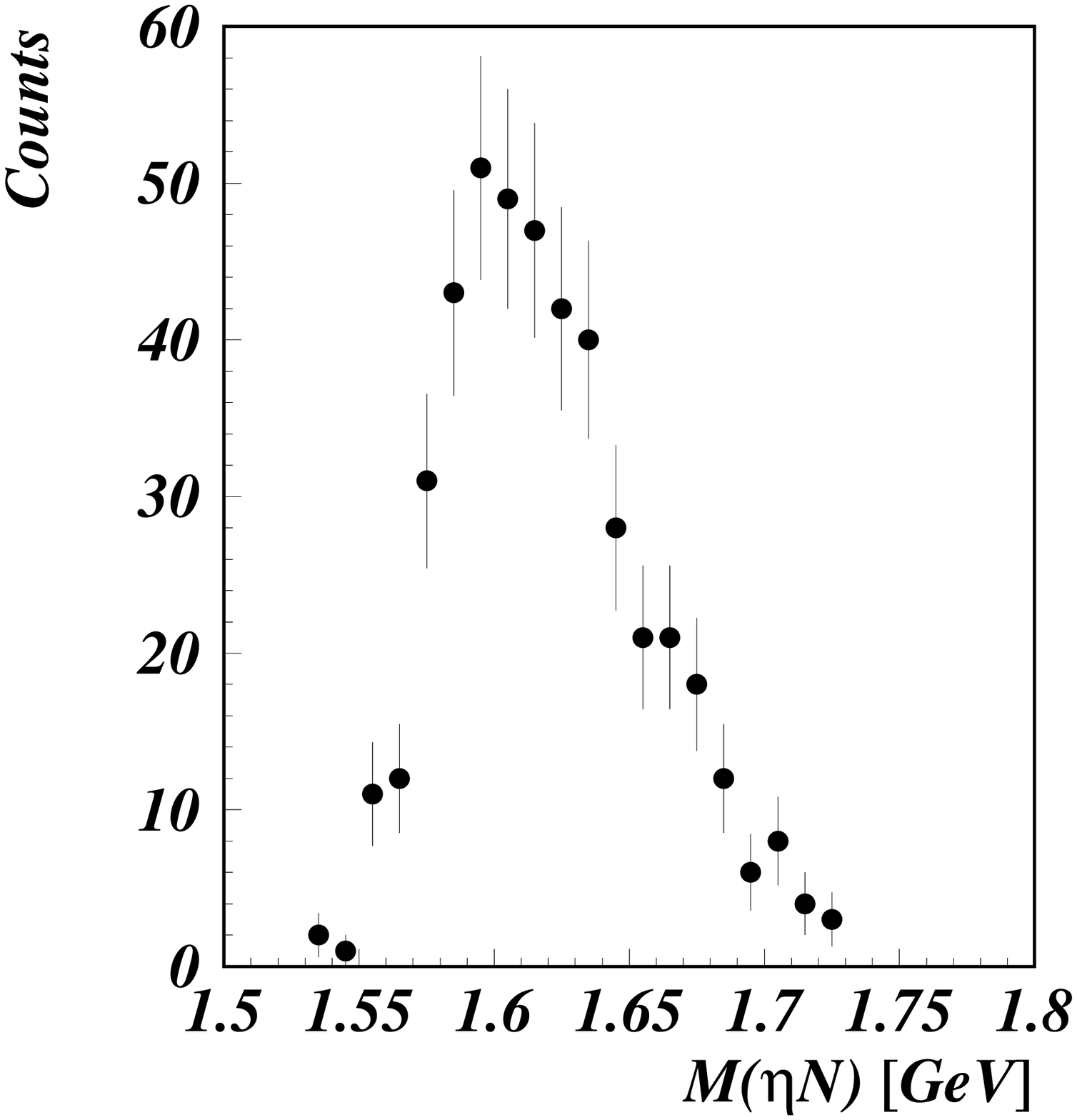}
%\protect\vspace*{-2.0 cm}
%\vspace*{2.0 cm}
\caption{Simulated spectra of $\eta N$ invariant masses for $\gamma p \to \eta \pi^0 p$
(on the left) and $\gamma p \to \eta \pi^+ n$ (on the right) reactions.}
\end{center} 
\vspace*{-0.5 cm}
\label{fig6}
\end{figure}

It is unclear if the interference of known wide resonances~\cite{ani} or
the cusp effect~\cite{dor} - two other hypotheses under discussion - could explain these results.

Our observation requires a confirmation from other groups. It would be
interesting to revisit the analysis of the $\gamma p \to \pi^0 \eta p$ reaction
by the CBELSA/TAPS Collaboration~\cite{gutz}.  If the similar energy binning
($\Delta E_{\gamma} = 1.2 - 1.45$ GeV) as in Ref.~\cite{gutz} is used in our
analysis and neither cut on $IM(\pi^0 p)$ is imposed, no signal of $N(1685)$ is
visible.  New dedicated experiments at other facilities could provide data
at a higher level of quality.

In summary, we report on the observation of narrow peak-like structure in the
$M(\eta p)$ and $M(\eta n)$ spectra in the $\gamma N \to \pi \eta N$
reactions. Quite likely these structures witnesses the existence of a new narrow
isospin-1/2 resonance $N(1685)$.

It is our pleasure to thank the staff of the European Synchrotron Radiation
Facility (Grenoble, France) for the stable beam operation during the
experimental runs.  This work was supported by High Energy Department of
Petersburg Nuclear Physics Institute and by the INFN Section of Catania,
and br the Russian Foundation of Basic Research grant 15-02-07873. 
Special thanks goes to Prof. E.~Gutz for
providing the data of the CBELSA/TAPS Collaboration and to Prof. M.~Oleynik for
the assistance in preparation of this manuscript.

\end{document}